\documentclass[a4paper,12pt]{article} 
\usepackage{amsmath,amsthm,amssymb}
\newtheorem{theorem}{Theorem}[section]

\newtheorem{proposition}[theorem]{Proposition}

\newcommand{\bm}[1]{\mbox{\boldmath$#1$}}
\newcommand{\be}{\begin{equation}}
\newcommand{\ee}{\end{equation}}
\newcommand{\bea}{\begin{eqnarray}}
\newcommand{\eea}{\end{eqnarray}}
\newcommand{\non}{\nonumber}
\begin{document}

\title{On the evaluation of form factors 
and correlation functions for the integrable spin-$s$ 
XXZ chains via the fusion method}
\author{Tetsuo Deguchi$^{1}$
\footnote{e-mail deguchi@phys.ocha.ac.jp} 
and Chihiro Matsui$^{2, 3}$
\footnote{e-mail matsui@spin.phys.s.u-tokyo.ac.jp}
}
\date{ }
\maketitle
\begin{center}  $^{1}$ 
Department of Physics, Graduate School of Humanities and Sciences, 
Ochanomizu University \\
2-1-1 Ohtsuka, Bunkyo-ku, Tokyo 112-8610, Japan 
\end{center} 
\begin{center} $^{2}$ 
Department of Physics, Graduate School of Science, the University of Tokyo \\ 
7-3-1 Hongo, Bunkyo-ku, Tokyo 113-0033, Japan \\ 
$^{3}$ CREST, JST, 4-1-8 Honcho Kawaguchi, Saitama, 332-0012, Japan
\end{center} 

\begin{abstract} 
Revising the derivation of the previous papers \cite{DM1,DM2,DM3}, 
 for the integrable spin-$s$ XXZ chain 
we express any form factor in terms of a single sum over scalar products 
of the spin-1/2 XXZ chain. With the revised method we express 
the spin-$s$ XXZ correlation function of 
any given entry at zero temperature 
in terms of a single sum of multiple integrals. 
\end{abstract}

\section{Form factors for the spin-$s$ XXZ spin chains}

Recently, a systematic method for evaluating the higher-spin form factors and correlation functions for the integrable spin-$s$ XXZ spin chain has been 
constructed through the fusion method \cite{DM1,DM2,DM3}. 
However, the method was not completely correct. There was a non-trivial 
assumption that the monodromy matrix should commute 
with the projection operator at an {\it arbitrary} rapidity. 
The transfer matrix may be non-regular or even singular 
if the rapidity is equal to one of inhomogeneous parameters forming complete strings, so that the quantum inverse-scattering formulas do not necessarily 
hold there. 
In this note we revise the method and show a formula by which we can express any spin-$s$ form factor in terms of scalar products of the spin-1/2 operators. We also revise the multiple-integral representation of the zero-temperature correlation function of an arbitrary entry, and express it in terms of a single sum of multiple integrals.  

Let us consider the spin-$\ell/2$ representation $V^{(\ell)}$ of the quantum group $U_q(sl_2)$ constructed in the $\ell$th tensor product $(V^{(1)})^{\otimes \ell}$  of the spin-1/2 representations $V^{(1)}$. 
The basis vectors $|| \ell, n \rangle$ derived from 
the highest weight vector: 
$|| \ell, 0 \rangle = | \uparrow \rangle_1 \otimes \cdots \otimes 
| \uparrow \rangle_{\ell}$ are given by  \cite{DM1}      
\begin{equation} 
|| \ell, n \rangle = 
\sum_{1 \le i_1 < \cdots < i_n \le \ell} 
\sigma_{i_1}^{-} \cdots \sigma_{i_n}^{-} || \ell, 0 \rangle \, 
q^{i_1+ i_2 + \cdots + i_n  - n \ell + n(n-1)/2} \quad  \mbox{for} \, \, 
n=0, 1, \ldots, \ell . 
\label{eq:|ell,n>} 
\end{equation}
Let $[n]_q$ denote the $q$-integer of an integer $n$: $[n]_q=(q^n-q^{-n})/(q-q^{-1})$, and $[n]_q!$ the $q$-factorial: $[n]_q!=\prod_{k=1}^{n} [k]_q$.  
We define the ``square length'' of $|| \ell, n \rangle$ by  
$F(\ell, n) = (|| \ell, n \rangle)^{t} \cdot || \ell, n \rangle = 
([\ell]_q ! /[\ell-n]_q! [n]_q!)  q^{-n(\ell-n)}.$ 
We define elementary matrices $E^{i, \, j \, (\ell \, p)}$ 
in the principal grading  by 
\be 
E^{i, \, j \, (\ell \, p)} = || \ell, i \rangle \, (|| \ell, j \rangle)^{t} 
/F(\ell, j) \quad 
\mbox{for} \, \, i, j = 0, 1, \ldots, \ell.  
\ee

We now construct the spin-$\ell/2$ monodromy matrix on the spin-$\ell/2$ chain 
with $N_s$ sites. Setting $L=\ell N_s$, we consider 
the $N_s$th tensor product $(V^{(\ell)})^{\otimes N_s}$ in 
$(V^{(1)})^{\otimes L}$. 
Let us denote by $\{ w_j \}_L$  a set of arbitrary parameters 
$w_{j}$ for $j=1 , 2, \ldots, L$, which we call inhomogeneous parameters. 
We define the spin-1/2 XXZ monodromy matrix by 
\be 
T^{(1)}(\lambda; \{ w_j \}_L)=R_{0, 1 2 \cdots L} 
 = R_{0 L}(\lambda-w_L) \cdots R_{0 1}(\lambda-w_1) \, . 
\ee
Here $R_{jk}(\lambda_j-\lambda_k)$ denote the symmetric $R$-matrices acting 
on the $j$th and the $k$th components of the tensor product space 
$(V^{(1)})^{\otimes L}$ where $\lambda_0=\lambda$ and $\lambda_j=w_j$ 
for $j=1, 2, \ldots, L$.  Let $\epsilon$ be 
infinitesimally small. We denote by $w_j^{(\ell; \, \epsilon)}$ 
the $N_s$ sets of almost complete $\ell$-strings \cite{DM2}:  
\be 
w_j^{(\ell; \, \epsilon)} = \xi_b - (\beta-1) \eta + \epsilon r_b^{(\beta)} 
\quad \mbox{ for} \, \, \beta= 1, 2, \ldots, \ell; \, b=1, 2, \ldots, N_s,  
\ee
where $r_b^{(\beta)}$ are generic. For $\epsilon=0$ we denote them by 
  $w_j^{(\ell)}$ and call them complete $\ell$-strings.   
We define $T^{(\ell; \, \epsilon)}(\lambda)$ by 
$T^{(\ell; \, \epsilon)}(\lambda) = T^{(1)}(\lambda; \{w_j^{(\ell; \, \epsilon)} \}_L)$ putting $w_j=w_j^{(\ell; \, \epsilon)}$ for $j=1, 2, \ldots, L$. 
Let us denote by $P^{(\ell)}_{j}$ the projector 
which maps the tensor product of the $j$th to the $(j+\ell-1)$th components of $(V^{(1)})^{\otimes L}$ onto the spin-$\ell/2$ representation 
$V^{(\ell)}$. 
We construct the spin-$\ell/2$ monodromy matrix $T^{(\ell)}(\lambda)$ 
by applying the projector $P^{(\ell)}_{1 \cdots L} := 
\prod_{b=1}^{N_s} P^{(\ell)}_{\ell (b-1)+1}$ as follows 
\cite{DM1} 
\be 
T^{(\ell)}(\lambda; \{ \xi_b \}_{N_s})= P^{(\ell)}_{1 \cdots L} 
T^{(\ell; \, 0)}(\lambda) P^{(\ell)}_{1 \cdots L} \, . 
\ee
Here we have defined $T^{(\ell; \, 0)}(\lambda)$ by the operator in the limit: $T^{(\ell; \, 0)}(\lambda) = \lim_{\epsilon \rightarrow 0} T^{(\ell; \, \epsilon)}(\lambda)$. 
We shall denote the $(\varepsilon, \varepsilon^{'})$-element of the spin-1/2 monodromy matrix $T^{(\ell; \, 0)}(\lambda)$ 
by $T^{(\ell; \, 0)}_{\varepsilon, \, \varepsilon^{'}}(\lambda)$, 
such as $T_{0, \, 1}^{(\ell; \, 0)} = B^{(\ell; \, 0)}(\lambda)$.  
We shall denote by $\{ \lambda_k \}_M$ a set of parameters 
$\lambda_k$ for $k=1, 2, \ldots, M$.

Let $|0 \rangle$ be the vacuum: 
$|0 \rangle = | \uparrow \rangle_1 \otimes \cdots 
\otimes | \uparrow \rangle_L$. 
We introduce variables $\varepsilon_{\alpha}^{'}$ and $\varepsilon_{\beta}$ 
which take only two values 0 and 1 for $\alpha, \beta = 1, 2, \ldots, \ell$. 
We define $e_j^{\varepsilon^{'}, \, \varepsilon}$ 
($\varepsilon^{'}, \varepsilon=0, 1$) 
by a two-by-two matrix which acts on the $j$th site 
 with only one nonzero element 1 at the entry of 
$(\varepsilon^{'}, \varepsilon)$ 
for each $j$ with $1 \le j \le \ell $. We shall define 
$\varepsilon_j^{'}$, $\varepsilon_j$ and $e_j^{\varepsilon^{'}, \, \varepsilon}$ also for $j$ satisfying $1 \le j \le L$, later. 
\begin{proposition} 
For arbitrary parameters $\{\mu_{k} \}_N$ and   
$\{\lambda_{\gamma} \}_M$ with $i_1-j_1=N-M$ 
we have 
\bea 
& & \langle 0 | \prod_{k=1}^{N} C^{(\ell)}(\mu_{k}) \, \cdot \, 
E^{i_1, \, j_1 \, (\ell \, p)}_1 \, \cdot \, \prod_{\gamma=1}^{M} 
B^{(\ell)}(\lambda_{\gamma}) | 0 \rangle 
= F(\ell, i_1)/F(\ell, j_1) \, \cdot \, 
q^{i_1(\ell-i_1)/2 - j_1(\ell-j_1)/2} 
\non \\ 
& & \qquad  \times \, 
\sum_{ \{ \varepsilon_{\beta} \} }  
\langle 0 | \prod_{k=1}^{N} C^{(\ell; \, 0)}(\mu_k) 
\, \cdot \, 
e_1^{\varepsilon_1^{'}, \varepsilon_1} \cdots 
e_{\ell}^{\varepsilon_{\ell}^{'}, \varepsilon_{\ell}} 
\, \cdot \, \prod_{\gamma=1}^{M} 
B^{(\ell; \, 0)}(\lambda_{\gamma}) | 0 \rangle \, .  
\label{eq:Ee}
\eea
Here we take the sum over all sets of 
$\varepsilon_{\beta}$ such that the number of integers ${\beta}$ 
satisfying $\varepsilon_{\beta}=1$ and $1 \le \beta \le \ell$ 
is given by $j_1$,  
while we take a set of $\varepsilon_{\alpha}^{'}$   
such that the number of integers ${\alpha}$ satisfying 
$\varepsilon_{\alpha}^{'}=1$ and $1 \le \alpha \le \ell$ 
is given by $i_1$. Each summand of (\ref{eq:Ee}) 
is symmetric with respect to exchange of $\varepsilon_{\alpha}^{'}$s: 
the following expression is independent of any 
permutation $\pi \in {\cal S}_{\ell}$  
\be 
\langle 0 | \prod_{k=1}^{N} C^{(\ell; \, 0)}(\mu_k) 
\, \cdot \, 
e_1^{\varepsilon_{\pi 1}^{'}, \, \varepsilon_1} \cdots 
e_{\ell}^{\varepsilon_{\pi \ell}^{'}, \, \varepsilon_{\ell}} 
\, \cdot \, \prod_{\gamma=1}^{M} 
B^{(\ell; \, 0)}(\lambda_{\gamma}) | 0 \rangle \, . 
\label{eq:sym-epsilon_a'}
\ee
\end{proposition} 
Here ${\cal S}_{n}$ denotes the set of permutations of 
$n$ integers, $1, 2, \ldots, n$.  

\begin{proposition} 
For a given set of the Bethe roots $\{ \lambda_{\gamma} \}_M$ 
we evaluate the scalar product (\ref{eq:sym-epsilon_a'}) through 
Slavnov's formula of scalar products for the spin-1/2 operators   
\bea  
& &  
\langle 0 | \prod_{k=1}^{N} C^{(\ell; \, 0)}(\mu_k) \, \cdot \, 
e_1^{\varepsilon_1^{'}, \varepsilon_1} \cdots 
e_{\ell}^{\varepsilon_{\ell}^{'}, \varepsilon_{\ell}} \, \cdot 
\, \prod_{\gamma=1}^{M} B^{(\ell; \, 0)}(\lambda_{\gamma}) | 0 \rangle \, 
 = \, \phi_{\ell}(\{ \lambda_{\gamma} \}; \{ w_j^{(\ell)} \}_L) \, \times \, 
 \nonumber \\ 
& & \times \,  \lim_{\epsilon \rightarrow 0} 
\langle 0 | \prod_{k=1}^{N} C^{(\ell; \, \epsilon)}(\mu_k) \, \cdot \, 
T^{(\ell; \, \epsilon)}_{\varepsilon_1, \varepsilon_1^{'}}
(w_1^{(\ell; \, \epsilon)}) \cdots 
T^{(\ell; \, \epsilon)}_{\varepsilon_{\ell}, \varepsilon_{\ell}^{'}}
(w_{\ell}^{(\ell; \, \epsilon)}) 
\, \cdot \, \prod_{\gamma=1}^{M} 
B^{(\ell; \, \epsilon)}(\lambda_{\gamma}(\epsilon)) | 0 \rangle  \, , 
\eea
where $\{ \lambda_{\gamma}(\epsilon) \}_M$ satisfy 
the spin-1/2 Beth ansatz equations with inhomogeneous parameters given 
by the almost complete $\ell$-strings: $w_j=w_j^{(\ell; \, \epsilon)}$ 
for $j=1, 2, \ldots, L$, 
and $\phi_{m}(\{ \lambda_{\gamma} \}; \{ w_j \}_L)$ has been defined by  
$\phi_{m} (\{ \lambda_{\gamma} \}; \{ w_j \}_L)= \prod_{j=1}^{m} 
\prod_{\gamma=1}^{M} b(\lambda_{\gamma}-w_j)$ 
with $b(u)=\sinh(u)/\sinh(u+\eta)$. 
\end{proposition}


For even $L$ we may assume that the ground state 
$| \psi_g^{(\ell; \, 0)} \rangle = 
\prod_{\gamma=1}^{M} B^{(\ell; \, 0)}(\lambda_{\gamma}) | 0 \rangle$ 
has the spin inversion symmetry: $U | \psi_g^{(\ell; \, 0)} \rangle = \pm | \psi_g^{(\ell; \, 0)} \rangle$ for $U=\prod_{j=1}^{L} \sigma_j^{x}$. 
We derive symmetry relations as follows.
\be 
\langle \psi_g^{(\ell; \, 0)} | \, 
e_1^{\varepsilon_1^{'}, \varepsilon_1} \cdots 
e_{\ell}^{\varepsilon_{\ell}^{'}, \varepsilon_{\ell}} 
\, 
| \psi_g^{(\ell; \, 0)} \rangle  
= \langle \psi_g^{(\ell; \, 0)} | \, 
e_1^{1- \varepsilon_1^{'}, \, 1 - \varepsilon_1} \cdots 
e_{\ell}^{ 1 - \varepsilon_{\ell}^{'}, \, 1 - \varepsilon_{\ell}} \, 
| \psi_g^{(\ell; \, 0)} \rangle \, . \label{eq:spin-inv}
\ee

\section{Spin-$s$ XXZ correlation functions in the massless regime}


Let us  now consider the spin-$s$ XXZ correlation functions,  
where $2s$ corresponds to integer $\ell$ 
of $V^{(\ell)}$. 
In the massless regime we set $\eta= i \zeta$ with $0 \le \zeta < \pi$. 
We assume that in the region $0 \le \zeta < \pi/2s$ 
the spin-$s$ ground state $| \psi_g^{(2s)} \rangle$ is 
given by $N_s/2$ sets of the $2s$-strings:   
\begin{equation} 
\lambda_{a}^{(\alpha)} 
= \mu_a - (\alpha- 1/2) \eta + \delta_a^{(\alpha)} \, , \quad  
\mbox{for} \, \, a=1, 2, \ldots, N_s/2 \, \,  
\mbox{and} \, \,  \alpha = 1, 2, \ldots, 2s .  
\end{equation} 
Here we also assume that string deviations $\delta_a^{(\alpha)}$ are  
small enough when $N_s$ is large enough. 
In terms of $\lambda_{a}^{(\alpha)}$, 
the spin-$s$ ground state associated with the principal grading 
is given by  
\begin{equation} 
 | \psi_g^{(2s)} \rangle = 
\prod_{a=1}^{N_s/2} \prod_{\alpha=1}^{2s} 
{B}^{(2s)}(\lambda_a^{(\alpha)}; \{\xi_b \}_{N_s}) | 0 \rangle . 
\end{equation}
Here we have $M$ Bethe roots with $M= 2s \, N_s/2 = s N_s$. Recall that 
$2s$ corresponds to $\ell$ of $V^{(\ell)}$.


We shall now formulate the multiple-integral representations 
of the spin-$s$ XXZ correlation functions for the most general case 
in the massless region: $0 \le \zeta < \pi/2s$.  
We define the zero-temperature correlation function 
for a given product of the spin-$s$ elementary matrices 
with principal grading 
 ${E}_1^{i_1 , \, j_1 \, (2s \, p)} \cdots 
{E}_m^{i_m, \, j_m \, (2s \, p)}$, which are   
$(2s+1) \times (2s+1)$ matrices, by  
\begin{equation}
F_m^{(2s)}(\{i_k, j_k \}) =  \langle \psi_g^{(2s)} | 
\prod_{k=1}^{m} {E}_k^{i_k , \, j_k \, (2s, \, p)} 
|\psi_g^{(2s)}  \rangle / \langle \psi_g^{(2s)} 
| \psi_g^{(2s)} \rangle \, . 
\label{eq:def-CF}
\end{equation}

For the $m$th product of the elementary matrices, 
we introduce sets of variables $\varepsilon_{\alpha}^{[k] \, '}$ 
and $\varepsilon_{\beta}^{[k]}$ ($1 \le k \le m$) such that 
the number of $\alpha$ satisfying $\varepsilon_{\alpha}^{[k] \, '}=1$ and 
$1 \le \alpha \le 2s$ is given by $i_k$ 
and the number of $\beta$ satisfying $\varepsilon_{\beta}^{[k]}=1$ and  
$1 \le \beta \le 2s$ by $j_k$, respectively. 
Here,  $\varepsilon_{\alpha}^{[k] \, '}$ 
and $\varepsilon_{\beta}^{[k]}$ take only two values 0 and 1. 
We express them also by variables $\varepsilon_j^{'}$ and $\varepsilon_j$ for 
$j=1, 2, \ldots, 2sm$ as  
\bea 
\varepsilon_{2s (k-1)+ \alpha}^{'}  & = & \varepsilon_{\alpha}^{[k] \, '} \quad \mbox{for} \quad \alpha=1, 2, \ldots, 2s;  k=1, 2, \ldots, m,  \nonumber \\  
\varepsilon_{2s (k-1)+ \beta} & = & \varepsilon_{\beta}^{[k]} \quad  
\mbox{for} \quad 
\beta=1, 2, \ldots, 2s;  k=1, 2, \ldots, m. 
\eea

For given sets of $\varepsilon_j$ and $\varepsilon_j^{'}$ for 
$j=1, 2, \ldots, 2sm$ we define 
$\mbox{\boldmath$\alpha$}^{-}$ by the set of integers $j$ 
satisfying $\varepsilon_j^{'}=1$ ($1 \le j \le 2sm$) 
and $\mbox{\boldmath$\alpha$}^{+}$ by 
the set of integers $j$ satisfying $\varepsilon_j=0$ ($1 \le j \le 2sm$):  
\begin{equation} 
\mbox{\boldmath$\alpha$}^{-}(\{ \varepsilon_j^{'} \}) 
= \{ j ; \, \varepsilon_j^{'}=1 \}  
\, , \quad 
\mbox{\boldmath$\alpha$}^{+}(\{ \varepsilon_j \}) 
= \{ j ; \, \varepsilon_j=0 \}  \, . 
\label{eq:def-aa'}
\end{equation} 
We denote by $r$ and $r^{'}$ the number 
of elements of the set $\mbox{\boldmath$\alpha$}^{-}$ 
and $\mbox{\boldmath$\alpha$}^{+}$, respectively. Due to charge conservation, 
we have $r + r^{'} = 2sm$. 
Precisely, we have $r= \sum_{k=1}^{m} i_k$ and 
$r^{'}= 2sm - \sum_{k=1}^{m} j_k$.     

For given sets ${\bm \alpha}^{-}$ and ${\bm \alpha}^{+}$, which correspond to 
$\{ \varepsilon_{j}^{'} \}_{2sm}$ and $\{ \varepsilon_{j} \}_{2sm}$, 
respectively,   
we define integral variables ${\tilde \lambda}_j$ for $j \in {\bm \alpha}^{-}$ 
and ${\tilde \lambda}^{'}_{j}$ for $j \in {\bm \alpha}^{+}$, 
respectively, by the following:   
\begin{equation} 
({\tilde \lambda}^{'}_{j^{'}_{max}}, \ldots, 
{\tilde \lambda}^{'}_{j^{'}_{min}},  {\tilde \lambda}_{j_{min}}, 
{\tilde \lambda}_{j_{max}})
=(\lambda_1, \ldots, \lambda_{2sm}) \, . 
\end{equation}

We now introduce a matrix 
$S=S\left( (\lambda_j)_{2sm}; (w_j^{(2s)})_{2sm} \right)$. 
For each integer $j$ satisfying $1 \le j \le 2sm$,  
we define $\alpha(\lambda_j)$ by $\alpha(\lambda_j)= \gamma$ 
with an integer $\gamma$ satisfying $1 \le \gamma \le 2s$  
if $\lambda_j$ is related to an integral variable $\mu_j$ through 
$\lambda_j = \mu_j - (\gamma - 1/2) \eta$ 
or if $\lambda_j$ takes a value close to $w_k^{(2s)}$ with $\beta(k)=\gamma$.
Thus, $\mu_j$ corresponds to the ``string center'' of $\lambda_j$. 
Here we have defined $\beta(j)$ by 
$\beta(j) = j - 2s [[(j-1)/2s]]$  
for $1 \le j \le M$. 
Here $[[x]]$ denotes the greatest integer less than or equal to $x$. 
We define the $(j,k)$ element of the matrix $S$ by   
\begin{equation} 
S_{j,k} = \rho(\lambda_j - w_k^{(2s)} + \eta/2) \, 
\delta(\alpha(\lambda_j), \beta(k)) \, , \quad {\rm for} \quad 
j, k= 1, 2, \ldots, 2sm \, .  
\end{equation} 
Here $\rho(\lambda)$ denotes the density of string centers \cite{DM2}, 
and $\delta(\alpha, \beta)$ the Kronecker delta.  
We obtain the following multiple-integral representation:     
\begin{eqnarray} 
& & F^{(2s)}_{m}(\{i_k, j_k\}) =  
\quad C(\{i_k, j_k \}) \, \times \nonumber \\ 
& & 
 \times \left( \int_{-\infty+ i \epsilon}^{\infty+ i \epsilon}
+ \cdots 
+ \int_{-\infty - i 
(2s-1) \zeta 
+ i \epsilon}
^{\infty - i 
(2s-1) \zeta  
+ i \epsilon} \right)  d \lambda_1 
\cdots 
\left( \int_{-\infty+ i \epsilon}^{\infty+ i \epsilon}
+ \cdots 
+ \int_{-\infty - i 
(2s-1) \zeta 
+ i \epsilon}^{\infty - i 
(2s-1) \zeta 
 + i \epsilon} 
\right)  d \lambda_{r^{'}} 
\nonumber \\  
& & \times \left( \int_{-\infty - i \epsilon}^{\infty - i \epsilon}
+ \cdots 
+ \int_{-\infty - i 
(2s-1) \zeta 
- i \epsilon}
^{\infty - i 
(2s-1) \zeta 
 - i \epsilon} \right)  
d \lambda_{r^{'} + 1} 
\cdots 
\left( \int_{-\infty - i \epsilon}^{\infty - i \epsilon}
+ \cdots 
+ \int_{-\infty - i 
(2s-1) \zeta 
 - i \epsilon}
^{\infty - i 
(2s-1) \zeta 
 - i \epsilon} 
\right)  d \lambda_{2sm}
\nonumber \\ 
& & \quad \times 
\sum_{{\bm \alpha}^{+}(\{ \epsilon_j \})} 
Q(\{ \varepsilon_j, \varepsilon_j^{'} \}; \lambda_1, \ldots, \lambda_{2sm}) \, 
{\rm det}
S(\lambda_1, \ldots, \lambda_{2sm}) \, . 
\label{eq:MIR}
\end{eqnarray}
Here the sum of ${\bm \alpha}^{+}(\{ \varepsilon_j \})$
is taken over all sets $\{ \varepsilon_j \}$ corresponding to 
$\{ \varepsilon_{\beta}^{[k]} \}$ $(1 \le k \le m)$ such that the number 
of integers $\beta$ satisfying $\varepsilon_{\beta}^{[k]}=1$ 
and $1 \le \beta \le 2s$ is given by $j_k$ for each $k$ satisfying 
$1 \le k \le m$.  
$Q(\{ \varepsilon_j, \varepsilon_j^{'} \}; \lambda_1, \ldots, \lambda_{2sm})$ 
is given by     
\begin{eqnarray} 
& & 
Q(\{ \varepsilon_j, \varepsilon_j^{'} \}; \lambda_1, \ldots, \lambda_{2sm})  
=(-1)^{r^{'}} \, 
{\frac { \prod_{j \in {\bm \alpha}^{-}( \{ \varepsilon_j^{'} \} ) }
\left( \prod_{k=1}^{j-1} 
\varphi({\tilde \lambda}_{j} - w_k^{(2s)} + \eta) 
\prod_{k=j+1}^{2sm} \varphi({\tilde \lambda}_{j} - w_k^{(2s)} ) \right)}
{\prod_{1 \le k < \ell \le 2sm} 
\varphi(\lambda_{\ell} - \lambda_{k} + \eta + \epsilon_{\ell, k})} } 
\nonumber \\ 
&& \qquad \qquad 
 \times \, \,   
 {\frac 
   {\prod_{j \in {\bm \alpha}^{+}( \{ \varepsilon_j \} ) } 
     \left( \prod_{k=1}^{j-1} 
            \varphi({\tilde \lambda}^{'}_{j} - w_k^{(2s)} - \eta) 
            \prod_{k=j+1}^{2sm} 
            \varphi({\tilde \lambda}^{'}_{j} - w_k^{(2s)} ) 
     \right) }
   {\prod_{1 \le k < \ell \le 2sm} 
     \varphi(w_{k}^{(2s)} - w_{\ell}^{(2s)}) } 
 } \, . \label{eq:Q}
\end{eqnarray}
In the denominator we set $\epsilon_{k, \ell}= i \epsilon$ for 
${Im}(\lambda_k -\lambda_{\ell}) > 0$ and 
$\epsilon_{k, \ell}= - i \epsilon$ for 
${Im}(\lambda_k -\lambda_{\ell}) < 0$,  where
$\epsilon$ is an infinitesimally small positive real number.  
The coefficient $C(\{i_k, j_k \})$ is given by 
\begin{equation} 
C(\{i_k, j_k \}) = 
\prod_{k=1}^{m} \left( F(\ell, i_k)/F(\ell, j_k) \, \cdot \, 
q^{i_k(\ell-i_k)/2-j_k(\ell-j_k)/2} \right) 
\, . \label{eq:Coef} 
\end{equation}
In (\ref{eq:Q}) we take a set ${\bm \alpha}^{-}(\{ \varepsilon_j^{'} \})$ 
corresponding to $\varepsilon_{\alpha}^{[k] \, '}$ 
for $k=1, 2, \ldots, m$, where the number of integers $\alpha$ satisfying 
$\varepsilon_{\alpha}^{[k] \, '}=1$ and $1 \le {\alpha} \le 2s$ 
is given by $i_k$ for each $k$ ($1 \le k \le m$).

We can derive the symmetric expression for the multiple-integral 
representation of the spin-$s$ correlation function 
$F^{(2s)}_{m}(\{i_k, j_k\})$ as follows. 
\begin{eqnarray}
&&  
F^{(2s)}_{m}(\{i_k, j_k\}) =  \frac{C(\{i_k, j_k \})}{\prod_{1 \leq \alpha < \beta \leq 2s}
\sinh^{m}(\beta-\alpha )\eta}     
\prod_{1\leq k < l \leq m}
\frac{\sinh^{2s}(\pi(\xi_k-\xi_l)/\zeta)}
{\prod^{2s}_{j=1}\prod^{2s}_{r=1}\sinh(\xi_k-\xi_l+(r-j)\eta)}   
\non \\
&&  \times \sum_{\sigma \in {\cal S}_{2sm}/({\cal S}_m)^{2s} } 
({\rm sgn} \, \sigma) \, 
 \prod^{r^{'}}_{j=1} 
\int^{\infty+ i \epsilon }_{-\infty + i \epsilon} d \mu_{\sigma j} \, 
\prod^{2sm}_{j=r^{'} +1} 
\int^{\infty - i \epsilon }_{-\infty - i \epsilon} d \mu_{\sigma j} 
\nonumber \\  
& &
%
\times \, \sum_{\{ \epsilon_{\beta}^{[1]} \} } \cdots 
\sum_{\{ \epsilon_{\beta}^{[m]} \} } \, 
Q^{'}(\{ \epsilon_j, \epsilon_j^{'} \}; \lambda_{\sigma 1}, \ldots, 
\lambda_{\sigma(2sm)})) \, 
\left( \prod^{2sm}_{j=1} 
{\frac {\prod^{m}_{b=1} \prod^{2s-1}_{\beta=1}
\sinh(\lambda_{j}-\xi_b+ \beta \eta)}
{\prod_{b=1}^{m} \cosh(\pi(\mu_{j}-\xi_b)/\zeta)}} \right) \non \\ 
& &\times \, {\frac {i^{2sm^2}} { (2 i \zeta)^{2sm} }} \, 
\prod^{2s}_{\gamma=1} 
\prod_{1 \le b < a \le m}
\sinh(\pi(\mu_{2s(a-1)+\gamma}-\mu_{2s(b-1)+\gamma})/\zeta) \, .  
 \label{eq:CFF2}
\end{eqnarray}
Here $\lambda_j$ are given by $\lambda_j= \mu_j - (\beta(j)-1/2) \eta$ for 
$j=1, \ldots, 2sm$, and (${\rm sgn} \, \sigma$) 
denotes the sign of permutation $\sigma \in {\cal S}_{2sm}/({\cal S}_m)^{2s}$. 
We have defined ${\cal S}_{2sm}/({\cal S}_m)^{2s}$ as follows \cite{DM2}:   
An element $\sigma$ of ${\cal S}_{2sm}/({\cal S}_m)^{2s}$ gives 
a permutation of integers $1, 2, \ldots, 2sm$, such that $\sigma j$ 
satisfying $\sigma j \equiv \beta$ (mod $2s$) are put in increasing order 
in the sequence $(\sigma 1, \sigma 2, \ldots, \sigma({2sm}) )$ 
for each integer $\beta$ satisfying $1 \le \beta \le 2s$.  
In (\ref{eq:CFF2}) we have defined $Q^{'}$ by 
multiplying $Q$ by the demoninator in the second line of (\ref{eq:Q}): 
 $Q^{'}(\{ \epsilon_j, \epsilon_j^{'} \}; \lambda_{1}, \ldots, \lambda_{2sm})
= Q(\{ \epsilon_j, \epsilon_j^{'} \}; \lambda_{1}, \ldots, \lambda_{2sm})
    {\prod_{1 \le k < \ell \le 2sm} 
     \varphi(w_{k}^{(2s)} - w_{\ell}^{(2s)}) }$.  
The coefficient $C(\{i_k, j_k \})$ is given by (\ref{eq:Coef}).  
The sums with respect to $\{ \varepsilon_{\beta}^{[k]} \}$ 
are taken over all sets $\{ \varepsilon_{\beta}^{[k]} \}$ $(1 \le k \le m)$ 
such that the number 
of integers $\beta$ satisfying $\varepsilon_{\beta}^{[k]}=1$ 
and $1 \le \beta \le 2s$ is given by $j_k$ for each $k$.  
We take such a set ${\bm \alpha}^{-}(\{ \varepsilon_j^{'} \})$ that 
corresponds to sets $\{ \varepsilon_{\alpha}^{[k] \, '}\}$ 
for $k=1, 2, \ldots, m$, where the number of integers $\alpha$ satisfying 
$\varepsilon_{\alpha}^{[k] \, '}=1$ and $1 \le {\alpha} \le 2s$ 
is given by $i_k$ for each $k$ ($1 \le k \le m$).

The spin-inversion symmetry (\ref{eq:spin-inv}) leads to useful 
relations among the expectation values of local or global operators. 
For an illustration, let us evaluate the one-point function 
in the spin-1 case with $i_1=j_1=1$, 
$\langle E_1^{1, \, 1 \, (2 \, p)} \rangle$. 
Setting $\varepsilon_1^{'}=0$ and $\varepsilon_2^{'}=1$ 
in formula (\ref{eq:Ee}) we decompose the 
spin-1 elementary matrix in terms of the spin-1/2 elementary matrices  
\be 
\langle \psi_g^{(2)} | E_1^{1, \, 1 \, (2 \, p)} | \psi_g^{(2)} \rangle 
= 
\langle \psi_g^{(2; \, 0)} | e_1^{0, \, 0} e_2^{1, \, 1} 
| \psi_g^{(2; \, 0)} \rangle
+ \langle \psi_g^{(2; \, 0)} | e_1^{0, \, 1} e_2^{1, \, 0} 
| \psi_g^{(2; \, 0)} \rangle \, .  
\ee
Through symmetry relations (\ref{eq:sym-epsilon_a'}) 
with respect to $\varepsilon_{\alpha}^{'}$  and the spin inversion 
(\ref{eq:spin-inv}) we have     
\be 
\langle \psi_g^{(2; \, 0)} | e^{0, \, 0}_{1}e_2^{1, \, 1} 
| \psi_g^{(2; \, 0)} \rangle  
= \langle \psi_g^{(2; \, 0)} | e^{1, \, 1}_{1}e_2^{0, \, 0} 
| \psi_g^{(2; \, 0)} \rangle 
=\langle \psi_g^{(2; \, 0)} | e^{0, \, 1}_{1}e_2^{1, \, 0} 
| \psi_g^{(2; \, 0)} \rangle  
= \langle \psi_g^{(2; \, 0)} | e^{1, \, 0}_{1} 
e_2^{0, \, 1} | \psi_g^{(2; \, 0)} \rangle \,,   
\ee 
and hence we have   
\be 
\langle \psi_g^{(2)} | E_1^{1, \, 1 \, (2 \, p)} | \psi_g^{(2)} \rangle 
= 2 \, \langle \psi_g^{(2; \, 0)} | e_1^{0, \, 0} e_2^{1, \, 1} 
| \psi_g^{(2; \, 0)} \rangle\, . 
\ee   
We thus obtain the double-integral representation 
of $\langle E_1^{1, \, 1 \, (2 \, p)} \rangle$ such as given in Ref. \cite{DM2}.

We would like to thank K. Motegi and J. Sato for valuable and helpful 
comments.

\end{document}